\documentclass[aps,pra,twocolumn,superscriptaddress]{revtex4-2}

\usepackage[english]{babel}
\usepackage{graphicx}
\usepackage{amsmath}
\usepackage{mathrsfs}
\begin{document}

\title{Chiral Nonlinear Optics and Optical Control}

\author{Cedric Dufresne}
\affiliation{Centre for Nanophotonics, Department of Physics,
Engineering Physics \& Astronomy, 64 Bader Lane,
Queen’s University, Kingston, Ontario, Canada K7L 3N6}
\email[Corresponding Author - Annabelle Makowski: ]{a.makowski@queensu.ca}
\author{Annabelle Makowski}
\affiliation{Centre for Nanophotonics, Department of Physics,
Engineering Physics \& Astronomy, 64 Bader Lane,
Queen’s University, Kingston, Ontario, Canada K7L 3N6}
\author{Nir Rotenberg}
\affiliation{Centre for Nanophotonics, Department of Physics,
Engineering Physics \& Astronomy, 64 Bader Lane,
Queen’s University, Kingston, Ontario, Canada K7L 3N6}

\begin{abstract}

Chiral light-matter interactions lie at the heart of emerging technologies such as quantum network protocols and quantum logic gates. In the few photon regime, it has been shown that chiral interactions between photons and a waveguide-embedded two-level quantum emitter can break reciprocity and impart a directional $\pi$ phase shift while the transmission remains intact. In this work, we present a model for multicolor, chiral nonlinear interactions in waveguides using a Green's Tensor formalism. We challenge previously held notions and demonstrate the complex photon dynamics hidden in multicolor light-matter interactions in the few photon regime. By modulating a stronger control beam, we can manipulate a weaker signal beam that contains much less than a single photon per emitter lifetime, on average. We develop equations for the transmission of the signal photons and removing the control photons to uncover the true strength of these nonlinearities, which we show is stronger than what is possible in symmetric geometries. The model predicts tunable unity extinction and up to 30\% amplification in the signal, a $\sim$100x increase from standard predictions in which control photons are present. We also predict a tunable 0-$\pi$ phase shift via control modulation with significant robustness to emitter imperfections.  Our model opens a new regime of directional nonlinear quantum light-matter interactions for study, providing a route to efficient all-optical control of photons.
\end{abstract}

\maketitle

\section{Introduction}
The rich physics describing the interaction of a single quantum emitter with light in a photonic waveguide, despite intense study in recent years, continues to yield new insights. 
For example, although the Maxwell's Equations which govern the propagation of light obey time-reversal symmetry, it is possible to couple a quantum emitter only to photons that propagate in a single direction. This effect, known now as \textit{chiral quantum optics} \cite{lodahl_chiral_2017}, is a consequence of the phase and vector nature of the near-field of a photonic mode and the transition dipole, and has been observed with both quantum dots \cite{sollner_deterministic_2015, coles_chirality_2016,mehrabad_chiral_2020} and atoms \cite{wang_realizing_2024, wang_tunable_2021} in systems ranging from waveguides to photonic resonators. Chiral quantum light-matter interactions enable non-reciprocal photonic elements such as circulators \cite{scheucher_quantum_2016, wang_tunable_2020}, isolators \cite{kawaguchi_optical_2021, sayrin_nanophotonic_2015,tang_-chip_2019,xia_reversible_2014} and quantum logic gates \cite{wang_high-performance_2023, zhang_chirality_2022,zhou_chiral_2021} and can be found at the core of proposals for quantum photonic circuits \cite{mccaw_reconfigurable_2024, sollner_deterministic_2015,xiao_chiral_2021} and networks \cite{mahmoodian_quantum_2016, wang_chiral_2022,ramos_non-markovian_2016}.

Similarly, although predictions of few-photon coherent nonlinearities with single quantum emitters date back to 1977 \cite{wu_observation_1977}, almost 40 years passed before they were observed, first with single QDs \cite{xu_coherent_2007} (albeit at ultralow efficiencies) and then with single organic molecules \cite{maser_few-photon_2016,turschmann_chip-based_2017}. More recently, strong nonlinearities were observed with average control photon number (per emitter lifetime)  $\bar{n}_{c}\lesssim1$ \cite{le_jeannic_dynamical_2022}, a direct consequence of the dramatic improvements to quantum light-matter interfaces \cite{lodahl_interfacing_2015}. Such nonlinearities enable all-optical control of quantum light states, and even the engineering of their very wavefunctions \cite{chang_quantum_2014}.

Here, we present a theory of coherent chiral nonlinear quantum optics, developing a model for these interactions from first principles. To do so, we build on earlier models of these nonlinear interactions \cite{maser_few-photon_2016}, which were developed for symmetrically-coupled emitters, but using a Green's Tensor formalism \cite{asenjo-garcia_atom-light_2017} that enables us to include chiral geometries. More specifically, we model the quantum light-matter interactions that arise when both control and signal continuous wave beams simultaneously (and efficiently) interact with a single quantum emitter, as shown in Fig.~1. We are then able to extract both the phase and amplitude of the scattered signal (and control) field, both of which differ greatly between the symmetric and chiral scenarios. Further, because we include dephasing, our model allows us to differentiate between the \textit{coherent} and \textit{total} scattering, an important distinction if such nonlinearities are to be used in emerging quantum technologies.

This paper is organized as follows: in Sec.~II we describe the system, in Sec.~III we explore Multicolor chiral QED (quantum electrodynamics) in waveguides, and in Sec.~IV we present equations and simulations for transmission, phase and the impact of imperfections before concluding in Sec.~V.

\section{The System}\label{sec:System}
\begin{figure*}[htbp]
    \centering
    \includegraphics[width=1.0\linewidth]{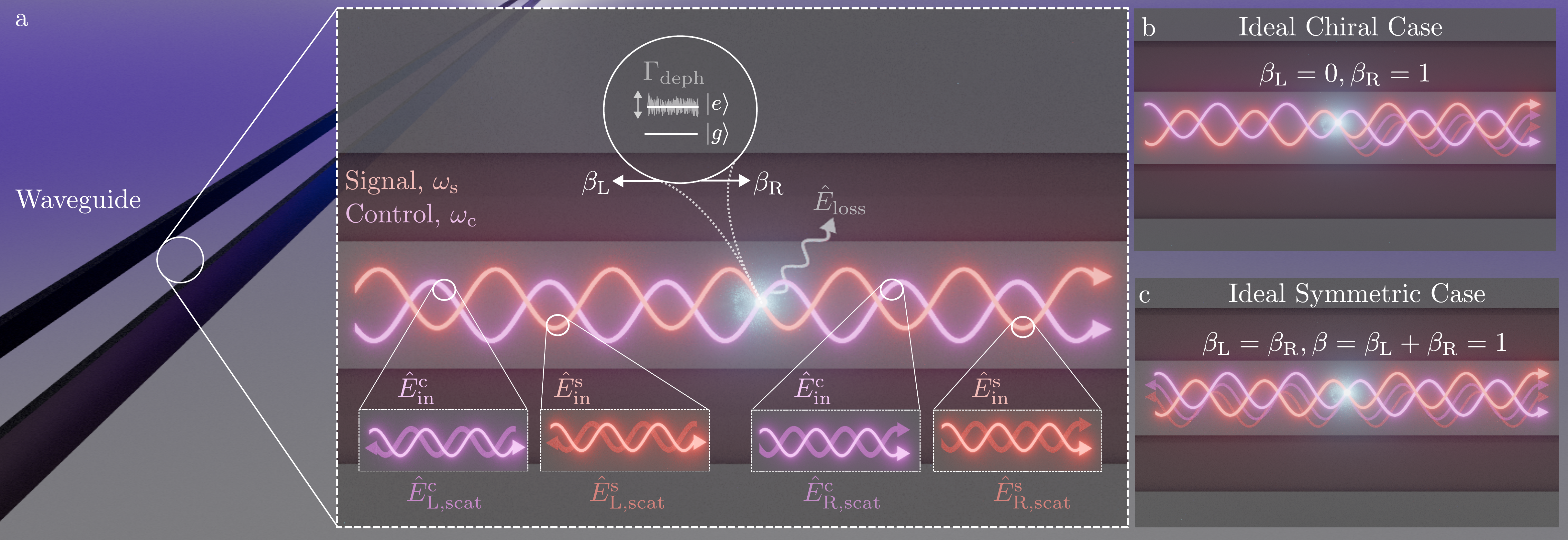}
    \caption{\textbf{a)} The TLE coupled to the waveguide interacting with the signal and control light fields. \textbf{b)} The ideal chiral case where only one mode will couple to the TLE, leading to directional emission. \textbf{c)} The ideal symmetric case where the TLE couples to both directions, leading to equal emission in either direction.}
    \label{fig:system_picture}
\end{figure*}

The system that we consider, which we sketch in Fig.~\ref{fig:system_picture}, is simply a two-level quantum emitter (TLE) coupled to a photonic waveguide that interacts with two light fields: the signal field, at frequency $\omega_\mathrm{s}$, and the control field at $\omega_\mathrm{c}$. In principle, these fields may enter or leave the waveguides in either end, but for clarity we consider the left end of the waveguide to be the input port, and we represent the input signal and control fields at any position $\mathbf{r}$ and time $t$ by their operators $\hat{\mathbf{E}}^{s}_{\mathrm{in}}\left(\mathbf{r}, t\right) \equiv\hat{\mathbf{E}}^{s}_{R,\, \mathrm{in}}\left(\mathbf{r}, t\right)$ and $\hat{\mathbf{E}}^{c}_{\mathrm{in}}\left(\mathbf{r}, t\right)\equiv\hat{\mathbf{E}}^{c}_{R,\, \mathrm{in}}\left(\mathbf{r}, t\right)$, where the first subscript denotes the direction in which the field propagates (but is dropped for clarity given that our input fields always travel to the right).

The input fields will propagate through the waveguide until they encounter, and possibly interact and scatter from the TLE. As sketched in Fig.~\ref{fig:system_picture}, the TLE has a transition frequency of $\omega_\mathrm{A}$ and associated decay rate $\Gamma$, and it couples to the left and right propagating modes with efficiencies $\beta_\mathrm{L}$ and $\beta_\mathrm{R}$, respectively. This can be expressed as, 
\begin{eqnarray}
    \beta_{\mathrm{L/R}} = \frac{\Gamma_{\mathrm{L/R}}}{\Gamma_{\mathrm{L}}+\Gamma_{\mathrm{R}}+\Gamma_{\mathrm{loss}}}
\end{eqnarray}
where $\Gamma_{\mathrm{loss}}$ is the rate of photons being scattered out of the waveguide via the TLE, $\Gamma_{\mathrm{L/R}}$ is the rate with which the emitter decays into the left- and right-propagating modes, respectively, and $\Gamma=\Gamma_{\mathrm{L}}+\Gamma_{\mathrm{R}}+\Gamma_{\mathrm{loss}}$. The complex geometries of photonic structures require numeric simulations by inserting a dipole within a structure to compute a $\beta$ value \cite{rotenberg_small_2017,javadi_numerical_2018}.
We can define the directionality of this coupling as,
\begin{equation}\label{eq:D}
    D=\frac{\beta_\mathrm{R}-\beta_\mathrm{L}}{\beta_\mathrm{R}+\beta_\mathrm{L}},
\end{equation}
where for ideal chiral coupling $D=1$ (the TLE only couples to photons traveling to the right) or $D=-1$ (the TLE only couples to photons traveling to the left), while for symmetric coupling $D=0$. We additionally consider a pure dephasing rate, $\Gamma_\mathrm{deph}$, corresponding to fast noise, typically due to phonons for solid-state emitters \cite{fan_pure_1998}.

Depending on the way that the TLE couples to the waveguide (i.e., the $\beta$'s), the output field differs. In Fig.~\ref{fig:system_picture} we show the two limiting cases: (b) for the ideal \textit{chiral} geometry where only one mode couples to the TLE (here, $\beta_\mathrm{L}=0$ and $\beta_\mathrm{R}=1$). In this case, all light exits the waveguide from the right end, with the output fields
\begin{subequations}
    \label{eq:EoutR}
    \begin{align}
        \hat{\mathbf{E}}^{s}_{R}\left(\mathbf{r}, t\right) &\equiv \hat{\mathbf{E}}^{s}_{R,\, \mathrm{out}}\left(\mathbf{r}, t\right) \nonumber \\
        &= \hat{\mathbf{E}}^{s}_{\mathrm{in}}\left(\mathbf{r}, t\right) + \hat{\mathbf{E}}^{s}_{R,\, \mathrm{scat}} \label{eq:EsRout}\left(\mathbf{r}, t\right), \\ 
        \hat{\mathbf{E}}^{c}_{R}\left(\mathbf{r}, t\right) &\equiv \hat{\mathbf{E}}^{c}_{R,\, \mathrm{out}}\left(\mathbf{r}, t\right) \nonumber \\ &= \hat{\mathbf{E}}^{c}_{\mathrm{in}}\left(\mathbf{r}, t\right) + \hat{\mathbf{E}}^{c}_{R,\, \mathrm{scat}}\left(\mathbf{r}, t\right),  \label{eq:EcRout}
    \end{align} 
\end{subequations}
being comprised of the input and scattered fields; (c) In the \textit{symmetric} geometry the TLE couples equally to both directions, meaning that $\beta_\mathrm{L}=\beta_\mathrm{R}$ (and, ideally, $\beta=\beta_\mathrm{L}+\beta_\mathrm{R}=1$) and backscattering by the TLE is possible. In this case, light also exits via the input port and in addition to Eq.~\ref{eq:EoutR} we find,
\begin{subequations}
    \label{eq:EoutL}
    \begin{align}
        \hat{\mathbf{E}}^{s}_{L}\left(\mathbf{r}, t\right) \equiv \hat{\mathbf{E}}^{s}_{L,\, \mathrm{out}}\left(\mathbf{r}, t\right) &= \hat{\mathbf{E}}^{s}_{L,\, \mathrm{scat}} \label{eq:EsLout}\left(\mathbf{r}, t\right), \\ 
        \hat{\mathbf{E}}^{c}_{L}\left(\mathbf{r}, t\right) \equiv  \hat{\mathbf{E}}^{c}_{L,\, \mathrm{out}}\left(\mathbf{r}, t\right) &= \hat{\mathbf{E}}^{c}_{L,\, \mathrm{scat}}\left(\mathbf{r}, t\right).  \label{eq:EcLout}
    \end{align} 
\end{subequations}
In each case, the total field can be decomposed into positive and negative frequency components. For example, $\hat{\mathbf{E}}^{s}_{L,\, \mathrm{scat}}=\hat{\mathbf{E}}^{s+}_{L,\, \mathrm{scat}} + \hat{\mathbf{E}}^{s-}_{L,\, \mathrm{scat}}$. 

Our main challenge is therefore to find expressions for the output field operators, and more specifically for the scattered field, in terms of the parameters of the TLE and input light fields. A secondary challenge that arises from our approach is that we calculate the total TLE population (i.e., the probability that it is in its excited state), but have no analytic expression for the fraction of this population due to the signal and control fields. As we shall see, in Sec. IV, we will have to develop a procedure to approximate these relative contributions if we want to, for example, calculate the transmission or reflection of just the signal (or control) fields.

\subsection{Observables}\label{sec:observables}
Once the field operators are known, we can use them to calculate measurable quantities of the quantum light fields, more specifically their amplitudes and phases. Starting with amplitude, we can calculate either the total transmission,
\begin{equation}\label{eq:Ttot}
    T=\frac{\left< \left(\tilde{\mathbf{E}}^{c-}_{\mathrm{in}}+\tilde{\mathbf{E}}^{s-}_{\mathrm{in}} + \tilde{\mathbf{E}}^-_{\mathrm{R,scat}}\right) \left( \tilde{\mathbf{E}}^{c+}_{\mathrm{in}}+\tilde{\mathbf{E}}^{s+}_{\mathrm{in}}+ \tilde{\mathbf{E}}^+_{\mathrm{R,scat}}\right)\right>}{\left< \left(\tilde{\mathbf{E}}^{c-}_{\mathrm{in}}+\tilde{\mathbf{E}}^{s-}_{\mathrm{in}}\right) \left( \tilde{\mathbf{E}}^{c+}_{\mathrm{in}}+\tilde{\mathbf{E}}^{s+}_{\mathrm{in}}\right)\right>},
\end{equation}
or the transmission of the signal and control fields, separately,
\begin{subequations}
    \label{eq:Tsc}
    \begin{align}
        T^s &= \frac{\left< \left( \tilde{\mathbf{E}}^{s-}_{\mathrm{in}} + \tilde{\mathbf{E}}^{s-}_{\mathrm{R,scat}}\right) \left( \tilde{\mathbf{E}}^{s+}_{\mathrm{in}} + \tilde{\mathbf{E}}^{s+}_{\mathrm{R,scat}}\right)\right>}{\left< \tilde{\mathbf{E}}^{s-}_{\mathrm{in}} \tilde{\mathbf{E}}^{s+}_{\mathrm{in}}\right>}, \label{eq:Ts} \\
        T^c &= \frac{\left< \left( \tilde{\mathbf{E}}^{c-}_{\mathrm{in}} + \tilde{\mathbf{E}}^{c-}_{\mathrm{R,scat}}\right) \left( \tilde{\mathbf{E}}^{c+}_{\mathrm{in}} + \tilde{\mathbf{E}}^{c+}_{\mathrm{R,scat}}\right)\right>}{\left< \tilde{\mathbf{E}}^{c-}_{\mathrm{in}} \tilde{\mathbf{E}}^{c+}_{\mathrm{in}}\right>},  \label{eq:Tc}
    \end{align}
\end{subequations}
where in these and the preceding equation, the spatial and temporal dependence is made implicit. Here, too, we have made the connection between theory and experiment explicit, noting that a measurement corresponds to a projection of the field operator onto a dipole $\mathbf{d}\left(\mathbf{r}\right)$ at the position of the measurement (and which represents the measurement device, for example a single photon detector). That is,
\begin{subequations}\label{eq:Etilda2hat}
    \begin{align}
        \tilde{\mathbf{E}}^{k+}_{j}\left(\mathbf{r}\right) &= \mathbf{d}^*\left(\mathbf{r}\right) \cdot \hat{\mathbf{E}}^{k+}_{j}\left(\mathbf{r}\right), \label{eq:Ep_tilda2hat} \\
        \tilde{\mathbf{E}}^{k-}_{j}\left(\mathbf{r}\right) &= \mathbf{d}\left(\mathbf{r}\right) \cdot \hat{\mathbf{E}}^{k-}_{j}\left(\mathbf{r}\right). \label{eq:Em_tilda2hat}
    \end{align}
\end{subequations}

In a similar manner, we can calculate the reflection at the input port,
\begin{equation}\label{eq:Rtot}
    R=\frac{\left< \tilde{\mathbf{E}}^-_{\mathrm{L,scat}} \tilde{\mathbf{E}}^+_{\mathrm{L,scat}}\right>}{\left< \left(\tilde{\mathbf{E}}^{c-}_{\mathrm{in}}+\tilde{\mathbf{E}}^{s-}_{\mathrm{in}}\right) \left( \tilde{\mathbf{E}}^{c+}_{\mathrm{in}}+\tilde{\mathbf{E}}^{s+}_{\mathrm{in}}\right)\right>},
\end{equation}
that in this case only depends on the field scattered to the right. This reflection, like the transmission, can be split into its signal and control components according to,
\begin{subequations}
    \label{eq:Rsc}
    \begin{align}
        R^s &= \frac{\left< \tilde{\mathbf{E}}^{s-}_{\mathrm{L,scat}}  \tilde{\mathbf{E}}^{s+}_{\mathrm{L,scat}}\right>}{\left< \tilde{\mathbf{E}}^{s-}_{\mathrm{in}} \tilde{\mathbf{E}}^{s+}_{\mathrm{in}}\right>}, \label{eq:Rs} \\
        R^c &= \frac{\left< \tilde{\mathbf{E}}^{c-}_{\mathrm{L,scat}} \tilde{\mathbf{E}}^{c+}_{\mathrm{L,scat}}\right>}{\left< \tilde{\mathbf{E}}^{c-}_{\mathrm{in}} \tilde{\mathbf{E}}^{c+}_{\mathrm{in}}\right>},  \label{eq:Rc}
    \end{align}
\end{subequations}
although to do this we must be able to split the total calculated scattered field into its constituent components.

Finally, instead of determining the intensity of the output field, we can calculate its phase. Because this phase is defined relative to the input field, and the signal and control fields may be of different frequencies, it makes no sense to attempt to define a phase of the total output field. Rather, we determine the phase of the output signal or control field by taking the \textit{argument} of the associated transmission coefficients,
\begin{subequations}
    \label{eq:tsc}
    \begin{align}
        t^s &= \frac{\left< \tilde{\mathbf{E}}^{s+}_{\mathrm{in}} + \tilde{\mathbf{E}}^{s+}_{\mathrm{R,scat}} \right>}{\left< \tilde{\mathbf{E}}^{s+}_{\mathrm{in}}\right>}, \label{eq:ts} \\
        t^c &= \frac{\left< \tilde{\mathbf{E}}^{c+}_{\mathrm{in}} + \tilde{\mathbf{E}}^{c+}_{\mathrm{R,scat}} \right>}{\left< \tilde{\mathbf{E}}^{c+}_{\mathrm{in}}\right>},  \label{eq:tc}
    \end{align}
\end{subequations}
or reflection coefficients,
\begin{subequations}
    \label{eq:rsc}
    \begin{align}
        r^s &= \frac{\left< \tilde{\mathbf{E}}^{s+}_{\mathrm{L,scat}} \right>}{\left< \tilde{\mathbf{E}}^{s+}_{\mathrm{in}}\right>}, \label{eq:rs} \\
        r^c &= \frac{\left< \tilde{\mathbf{E}}^{c+}_{\mathrm{L,scat}} \right>}{\left< \tilde{\mathbf{E}}^{c+}_{\mathrm{in}}\right>}.  \label{eq:rc}
    \end{align}
\end{subequations}
These, as we shall see in Sec.~IV, in conjunction with the transmission and reflection, also allow us to separate the \textit{coherent} and \textit{incoherent} contributions to the output fields \cite{mccaw_reconfigurable_2024}; in the presence of noise such as dephasing, $T\neq\left|t\right|^2$ and $R\neq\left|r\right|^2$.

\section{Multicolor Chiral QED in Waveguides}
The multicolor quantum light-matter interactions that we are interested in, shown in Fig.~1, can be described by the standard Hamiltonian \cite{cohentannoudji_atomphoton_1993}, expanded to explicitly include both the signal and control fields, and the different directions in which they travel. By working in a frame that rotates at $\omega_c$, and assuming a single mode (that can travel in both directions), monochromatic drive fields (at $\omega_s$ and $\omega_c$) and making the rotating wave approximation, we can write this Hamiltonian as,
\begin{widetext}
\begin{eqnarray}
    \hat{H} = &-&\hbar\Delta\hat{\sigma}_{eg}\hat{\sigma}_{ge} 
     + \hbar\omega_s\int d\mathbf{r} \, \left(\hat{\mathbf{f}}_R^{s\dagger}\left(\mathbf{r}\right) \hat{\mathbf{f}}^s_R\left(\mathbf{r}\right)+ \hat{\mathbf{f}}_L^{s\dagger}\left(\mathbf{r}\right) \hat{\mathbf{f}}^s_L\left(\mathbf{r}\right)\right)
     + \hbar\omega_c \int d\mathbf{r} \, \left(\hat{\mathbf{f}}_R^{c\dagger}\left(\mathbf{r}\right) \hat{\mathbf{f}}^c_R\left(\mathbf{r}\right)+ \hat{\mathbf{f}}_L^{c\dagger}\left(\mathbf{r}\right) \hat{\mathbf{f}}^c_L\left(\mathbf{r}\right)\right) \nonumber \\ 
     &-& \left(\mathbf{d}^{c*}_\mathrm{R}\cdot\hat{\mathbf{E}}^{c+}_{\mathrm{R}}+\mathbf{d}^{c*}_\mathrm{L}\cdot\hat{\mathbf{E}}^{c+}_{\mathrm{L}}\right) \hat{\sigma}_{eg} 
     - \left( \mathbf{d}^{c}_\mathrm{R}\cdot\hat{\mathbf{E}}^{c-}_{\mathrm{R}}+\mathbf{d}^{c}_\mathrm{L}\cdot\hat{\mathbf{E}}^{c-}_{\mathrm{L}}\right)\hat{\sigma}_{ge} \label{eq:H} \\
     &-& \left(\mathbf{d}^{s*}_\mathrm{R}\cdot\hat{\mathbf{E}}^{s+}_{\mathrm{R}}+\mathbf{d}^{s*}_\mathrm{L}\cdot\hat{\mathbf{E}}^{s+}_{\mathrm{L}}\right)\hat{\sigma}_{eg}e^{i\delta t}
     -  \left(\mathbf{d}^{s}_\mathrm{R}\cdot\hat{\mathbf{E}}^{s-}_{\mathrm{R}}+\mathbf{d}^{s}_\mathrm{L}\cdot\hat{\mathbf{E}}^{s-}_{\mathrm{L}}\right)\hat{\sigma}_{ge}e^{-i\delta t}. \nonumber
\end{eqnarray}    
\end{widetext}

 Here, the detuning $\Delta = \omega_c-\omega_A$, $\hat{\sigma}_{ij} = |i\rangle\langle j|$ are the atomic operators,  $\textbf{d}^\ell_d$ are the transition dipole moments and $\delta = \omega_s - \omega_c$ is the detuning between the signal and control fields. Finally, $\hat{f}_d^{\ell\dagger}\left(\mathbf{r}\right)$ and $\hat{f}_d^{\ell}\left(\mathbf{r}\right)$ are the spatially-dependent creation and annihilation operators for electromagnetic excitations in dispersive and/or absorbing media \cite{gruner_green-function_1996}, which obey the usual commutation relation,
\begin{equation}\label{eq:fCom}
    \left[\hat{\textbf{f}}^i_j(\mathbf{r},\omega),\hat{\textbf{f}}^{i'\dagger}_{j'}(\mathbf{r}',\omega')\right] = \delta_{i,i'}\delta_{j,j'}\delta(\mathbf{r}-\mathbf{r}')\delta(\omega-\omega').
\end{equation}

Finally, we note that Eq.~\ref{eq:H} describes the general scenario, where incident fields can travel from both the left and right. For the scenarios we consider, which we sketch in Fig.~1a, all incident fields travel from the left to the right, and therefore in what follows we take $\hat{\mathbf{E}}^{s}_{\mathrm{L,in}}=\hat{\mathbf{E}}^{c}_{\mathrm{L,in}}=0$.

\subsection{Emitter Dynamics}
In the presence of two fields of differing frequencies, the total field beats in time and hence so too does the response of an emitter in this field. To calculate this dynamic response, we begin by solving the von Neumann equation \cite{meystre_interaction_1990},
\begin{equation}\label{eq:vonNeumann}
    i\hbar \frac{d\hat{\rho}}{dt} = [\hat{H},\hat{\rho}] +\mathscr{L},
\end{equation}
for the TLE density matrix $\hat{\rho}$, using the Lindblad superoperator \cite{lindblad_generators_1976},
\begin{equation}\label{eq:L}
    \mathscr{L}=\sum\frac{\Gamma_{ji}}{2}\left(2\sigma_{ij}\rho\sigma_{ji}-\sigma_{jj}\rho-\rho\sigma_{jj}\right).
\end{equation}
Here, $i$ and $j$ range over the states $|g\rangle$ and $|e\rangle$, and $\Gamma_{eg}=\Gamma$ is the TLE spontaneous emission rate and $\Gamma_{gg}=\Gamma_{ee}=\Gamma_{\mathrm{deph}}$ is its pure dephasing rate. We solve Eq.~\ref{eq:vonNeumann} using the Hamiltonian of Eq.~\ref{eq:H}, finding that,
\begin{eqnarray}
    \dot{\rho}_{ee} &=& -\Gamma\rho_{ee} +i \rho^*_{ge}\left(\Omega^c_{R}+\Omega^s_{R}e^{i\delta t}\right) \nonumber \\ 
    && -i\rho_{ge}\left(\Omega^c_{R}+\Omega^s_{R}e^{-i\delta t}\right),  \label{eq:drho_ee}\\
    \dot{\rho}_{ge} &=& -\left(\Gamma_2 +i\Delta\right)\rho_{ge} +i\left(\Omega^c_{R}+\Omega^s_{R}e^{i\delta t}\right)\left(1-\rho_{ee}\right) \nonumber \\
    &&-i\left(\Omega^c_{R}+\Omega^s_{R}e^{i\delta t}\right)\rho_{ee}, \label{eq:drho_ge}
\end{eqnarray}

where $\Gamma_2=\Gamma/{2}+\Gamma_\mathrm{deph}$, and we assume coherent drive fields, allowing us to write $\Omega^\ell_d = \left<\hat{\textbf{d}}_d^{\ell*} \cdot \hat{E}_d^\ell \right>  / \hbar$ which we take to be real valued. Note that these two equations are sufficient to fully determine the entire density matrix because $\rho_{eg} = \rho^*_{ge}$ and $\rho_{gg} = 1-\rho_{ee}$.

Recall that these equations describe the case where all light is incident from the left; to recover the general scenario, where the incident beams may travel in both directions, simply replace $\Omega^s_{R} \rightarrow\Omega^s_{R}+\Omega^s_{L}$ and $\Omega^c_{R} \rightarrow\Omega^c_{R}+\Omega^c_{L}$.

While Eqs.~\ref{eq:drho_ee} and~\ref{eq:drho_ge} are exact, the steady-state response of the TLE cannot be solved for analytically, due to the oscillatory terms in each. Rather, we employ a Bloch vector approach \cite{jelezko_pumpprobe_1997,allen_optical_2012,cohentannoudji_optical_1998}, writing,
\begin{subequations}\label{eq:uvw}
    \begin{align}
        u &= \frac{1}{2}(\rho_{eg}+\rho_{ge}), \label{eq:u} \\
        v&=\frac{1}{2i}(\rho_{eg}-\rho_{ge}), \label{eq:v}\\
        w&=\frac{1}{2}(\rho_{ee}-\rho_{gg}), \label{eq:w}
    \end{align}
\end{subequations}
which correspond to the real and imaginary components of the coherence, and the population inversion, respectively. We put this together to create the Bloch vector, $\mathbf{x}=\left(\begin{array}{ccc}
u & v & w\end{array}\right)^{\mathrm{T}}$. We rewrite Eqs.~\ref{eq:drho_ee} and \ref{eq:drho_ge} in this basis to obtain the concise equation for the emitter dynamics,
\begin{equation}\label{eq:dx}
    \boldsymbol{\dot{\mathbf{x}}}=\left(\mathbf{A}-\Omega_{R}^{s}e^{i\delta t}\mathbf{B}-\Omega_{R}^{s}e^{-i\delta t}\mathbf{B}^{*}\right)+\mathbf{y},
\end{equation}
where,
\begin{equation}\label{eq:AB}
\mathbf{A} =\begin{pmatrix} -\Gamma_{2} & \Delta & 0\\
-\Delta & -\Gamma_{2} & 2\Omega_{R}^{c}\\
0 & -2\Omega_{R}^{c} & \Gamma
\end{pmatrix}, \quad \mathbf{B}=\begin{pmatrix}0 & 0 & -i\\
0 & 0 & -1\\
i & 1 & 0
\end{pmatrix},
\end{equation}
and
\begin{equation}\label{eq:y}
    \mathbf{y}=\left(\begin{array}{ccc}
0 & 0 & -\Gamma/2\end{array}\right)^{\mathrm{T}},
\end{equation}
where we again remind the reader that to recover the more general scenario, where incident fields can enter from both sides of the waveguide, we simply replace $\Omega^s_{R} \rightarrow\Omega^s_{R}+\Omega^s_{L}$ and $\Omega^c_{R} \rightarrow\Omega^c_{R}+\Omega^c_{L}$ (and if, for example, the signal field initially travels to the right while the control initially travels to the left then only replace $\Omega^c_{R} \rightarrow\Omega^c_{L}$). 

To solve Eqs.~\ref{eq:dx} to~\ref{eq:y} we use a Fourier Ansatz, assuming that $\mathbf{x}$ can be expanded in a Fourier series with harmonics that are multiples of the beat frequency $\delta$,
\begin{equation}\label{eq:xn}
    \textbf{x}=\sum_{n=0}^\infty \mathbf{x}_{n}e^{in\delta t}.
\end{equation}
This results in an infinite recursive relation
\begin{equation}
    -\mathbf{y}\delta\left(n,0\right)=(\mathbf{A}-in\delta)\mathbf{x}_{n}-\mathbf{B}\mathbf{x}_{n-1}-\mathbf{B}^{*}\mathbf{x}_{n+1}. 
\end{equation}
Typically, we need only solve this equation for the $n$ up to about 10 (see Appendix A) to converge such that truncation errors are below $10^{-14}$. Once $\mathbf{x}$ is known, we can invert Eqs.~\ref{eq:uvw} to recover the density matrix elements,
\begin{subequations}\label{eq:rhos}
    \begin{align}
        \rho_{ee} &= \sum_n\left[w_ne^{in\delta t} + \frac{1}{2}\delta\left(n,0\right)\right], \label{eq:rho_ee_n} \\
        \rho_{ge}&= \sum_n \left(u_n + iv_n\right)e^{in\delta t}. \label{eq:rho_ge_n}
    \end{align}
\end{subequations}
While these equations allow us to calculate the dynamics of the TLE population and coherences, in an experiment we would typically measure to time-averaged response. From the above equations, we can directly see that.
\begin{subequations}\label{eq:TArho}
    \begin{align}
        \left<\rho_{ee}\right> &= w_0 + \frac{1}{2}, \label{eq:TArho_ee} \\
        \left<\rho_{ge}^s\right>&=u_1 + iv_1, \label{eq:TArho_ge_s} \\
        \left<\rho_{ge}^c\right>&=u_0 + iv_0, \label{eq:TArho_ge_c}.
    \end{align}
\end{subequations}

\subsection{The light fields}
Having found the emitter response to the signal and control light fields, we now turn to the fields themselves. Because the different fields in our system commute (c.f. Eq.~\ref{eq:fCom}), we can consider them individually, each time, using the Von Neummann equation. Doing so results in a concise expression for the total scattered field (see Appendix B) \cite{maser_few-photon_2016,dung_resonant_2002},
\begin{equation} \label{eq:Escat_total_Green}
    \hat{\boldsymbol{E}}_{d,\mathrm{ scat}}^{+}\left(\mathbf{r},t\right)=\mu_{o}\omega_\mathrm{A}^{2}\mathbf{G}_d(\mathbf{r},\mathbf{r}_\mathrm{A},\omega_\mathrm{A})\cdot\mathbf{d}\hat{\boldsymbol{\sigma}}_{ge}(t),
\end{equation}
and for the components of the scattered field,
\begin{equation} \label{eq:Escat_sc_Green}
    \hat{\boldsymbol{E}}_{d,\mathrm{ scat}}^{\ell+}\left(\mathbf{r},t\right)=\mu_{o}\omega_\mathrm{A}^{2}\mathbf{G}_d(\mathbf{r},\mathbf{r}_\mathrm{A},\omega_\mathrm{A})\cdot\mathbf{d}\hat{\boldsymbol{\sigma}}_{ge}^\ell(t),
\end{equation}
where, recall, $d=\mathrm{L,R}$ and $\ell=s,c$.

It is typically more convenient to work with the dipole-projected Green tensor,
\begin{eqnarray}
    g_{ij,\mathrm{d}}(\mathbf{r}_i,\mathbf{r}_j,\omega) &=& \frac{\mu_0 \omega^2}{\hbar}\textbf{d}^*\left(\mathbf{r}_i\right) \cdot \textbf{G}_\mathrm{d}\left(\mathbf{r}_i,\mathbf{r}_j,\omega\right)\cdot \textbf{d}\left(\mathbf{r}_j\right), \nonumber\\
    &=& i\Gamma\beta_{\mathrm{d}}(\omega)e^{ik|\mathbf{r}_i-\mathbf{r}_j|}, \label{eq:g_ij}
\end{eqnarray}
where the second line is for a one-dimensional waveguide \cite{asenjo-garcia_atom-light_2017} and we assume that the emitter frequency already accounts for the Lamb Shift. Note, too, that while $\beta_{\mathrm{d}}\left(\omega\right)$ depends on the photon frequency, it typically constant over the small frequency ranges that we consider (i.e., $\beta_{\mathrm{d}}(\omega_s)\approx\beta_{\mathrm{d}}(\omega_c)\equiv\beta_{\mathrm{d}}$).  

In light of Eq.~\ref{eq:g_ij}, and keeping in mind the observables defined in Sec.~\ref{sec:observables}, we can rewrite Eqs.~\ref{eq:Escat_total_Green} and~\ref{eq:Escat_sc_Green} as,
\begin{eqnarray}
    \tilde{\boldsymbol{E}}_{d,\mathrm{ scat}}^{+}\left(\mathbf{r},t\right) &=& i\hbar \Gamma\beta_{\mathrm{d}}  \hat{\boldsymbol{\sigma}}_{ge}(t) e^{ik|\mathbf{r}-\mathbf{r}_\mathrm{A}|}, \label{eq:Escat_total} \\
    \tilde{\boldsymbol{E}}_{d,\mathrm{ scat}}^{\ell +}\left(\mathbf{r},t\right) &=& i\hbar \Gamma\beta_{\mathrm{d}}  \hat{\boldsymbol{\sigma}}_{ge}^\ell(t) e^{ik|\mathbf{r}-\mathbf{r}_\mathrm{A}|}, \label{eq:Escat_sc}
\end{eqnarray}
where for $\mathrm{d}=\mathrm{R}$, $\mathbf{r}>\mathbf{r}_\mathrm{A}$ and for $\mathrm{d}=\mathrm{L}$, $\mathbf{r}<\mathbf{r}_\mathrm{A}$.

\section{Results and Discussion}\label{sec:results}
\subsection{Total Transmission and Reflection}\label{sec:TRtot}
We are now ready to investigate the directional scattering of multiple light beams of differing frequencies by a quantum emitter coupled to a photonic waveguide. We begin by considering the transmission and reflection of the total light field, which contains both the signal and control components. We do so by using Eq.~\ref{eq:Escat_total} in Eqs.~\ref{eq:Ttot} and~\ref{eq:Rtot}, resulting in (see Appendix C for derivation),
\begin{eqnarray}
    T&=&1 - \frac{2\Gamma\beta_\mathrm{R}\mathrm{Im}\left\{\Omega_\mathrm{R}^s \rho_{eg,1} + \Omega_\mathrm{R}^c \rho_{eg,0}\right\} - \left(\Gamma\beta_\mathrm{R}\right)^2\left|\rho_{ee,0}\right|}{\left(\Omega_\mathrm{R}^s\right)^2 + \left(\Omega_\mathrm{R}^c\right)^2}, \label{eq:T}\\
    R&=&\frac{\Gamma^2\beta_\mathrm{L}\beta_\mathrm{R}\left| \rho_{ee,0}\right|}{\left(\Omega_\mathrm{R}^s\right)^2 + \left(\Omega_\mathrm{R}^c\right)^2}. \label{eq:R}
\end{eqnarray}

\begin{figure}[!tbhp]
    \centering
    \includegraphics[width=\linewidth]{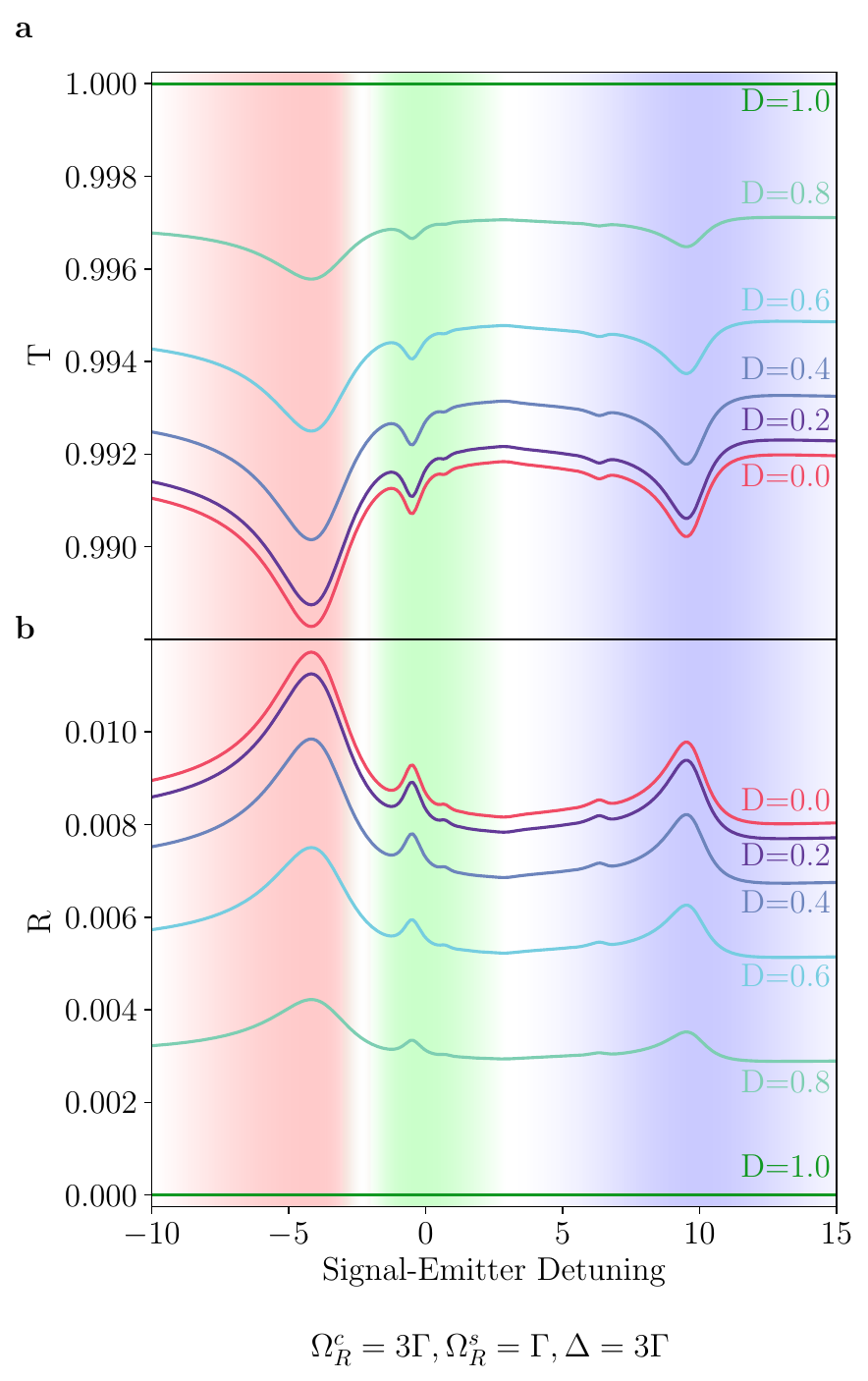}
    \caption{The transmission and reflection spectra when both the signal and control beams are interacting with the TLE. This is considering a perfect system with unity coupling and assuming no scattering losses or noise. Signatures of the TLE dressed state interaction can be observed at different signal-emitter detunings: three photon amplification is shown in red, energy transfer between the signal and control beams is shown in green, and the AC-Stark effect is shown in blue. These effects are discussed in more detail in Sec.~IV B below.}
    \label{fig:tot_trans_ref}
\end{figure}

We begin by considering an ideal system, that is one with unity coupling (i.e., $\beta=1$) so that there are no scattering losses, and no noise (i.e., $\Gamma_\mathrm{deph}=0$), adding imperfections in Sec.~IV D below. We present that exemplary $T$ and $R$ spectra for this idealized scenario in Fig.~\ref{fig:tot_trans_ref}, setting $\Delta=3\Gamma$ and working at $\Omega_\mathrm{R}^s=\Gamma$. This latter condition ensures that the interaction between the TLE and signal light field remains coherent and that this signal does not induce any power broadening \cite{c_cohen-tannoudji_dressed_1998}. 

The dependence of the $T$ spectrum on the directionality is presented in Fig.~\ref{fig:tot_trans_ref}a, here for a moderate $\Omega_\mathrm{R}^c=3\Gamma$. As expected, for perfectly chiral coupling $\left(D=1\right)$, $T=1$ for all $\omega_s$; this remains true for all values of $\Omega_\mathrm{R}^c$ and likewise for $D=-1$. For this chiral coupling, there is no corresponding reflection as shown in Fig.~\ref{fig:tot_trans_ref}b. As $D$ decreases towards 0 we begin to observe a signature of the interactions with the TLE. If the control drive field is increased, then we see a dampening of the nonlinear TLE interactions.

% Total simplified transmission and reflection figure

\subsection{Controlling the Signal Intensity}\label{sec:Tsig}
Often, instead of the total transmission and reflection, we are more interested in just the signal field. This could be because the signal field effectively probes the state of the dressed TLE \cite{mollow_stimulated_1972} meaning that it allows us to probe the way a light field can control a quantum emitter, or for quantum of ultra-low power optical technologies. To remind the reader, the problem is that in Eq.~\ref{eq:Ts} (and in light of Eqs.~\ref{eq:Escat_sc_Green} and \ref{eq:Escat_sc}), the term $\left<\sigma_{eg}^s \sigma_{ge}^s\right> = \left<\sigma_{ee}^s \right> = \rho_{ee}^s$ arises, yet we only calculate the total population $\rho_{ee}$. There are different ways to address this issue, which depend on what exactly one wishes to measure and, in fact, we can envision 3 different ways in which we may isolate and measure the signal field, which we derive expressions for.

\subsubsection{Suppressed control beam}
First, one may \textit{suppress the control} 
for example by changing the angle with which it is incident in free space \cite{maser_few-photon_2016} or, in a waveguide, through free-space excitation \cite{turschmann_chip-based_2017}, by having it counter-propagate relative to the signal field \cite{le_jeannic_dynamical_2022} or potentially by using a different mode. Regardless of the method used, and even for perfect nanophotonic systems (i.e., those with no unwanted scattering for example due to surface roughness), the TLE itself will scatter (or emit) some of the control photons into the signal path. This unwanted scattering is equivalent to the reflection of just the control beam (c.f. Eq.~\ref{eq:Rc}) as just with the reflection only the scattered field is present (and not the incident), and its amplitude is therefore,
\begin{equation}
    A_{c\rightarrow s}^{\mathrm{T}} =\frac{(\beta_\mathrm{R}\Gamma)^{2}\rho_{ee}^{c}}{\left(\Omega_{\mathrm{R}}^{s}\right)^{2}},
\end{equation}
 where $\rho_{ee}^{c}$ is the emitter population when the signal field is either off or far detuned (i.e. does not interact with the TLE). This allows us to renormalize the transmitted signal (Eq.~\ref{eq:Ts}) to account for the scattered pump photons,
\begin{equation}\label{eq:TsNorm}
    T^s_{\mathrm{norm}}= \dfrac{1-2\dfrac{\beta_\mathrm{R}\Gamma}{\Omega_{\mathrm{R}}^{s}} \mathrm{Im}\left\{\rho_{eg,1}\right\}+ \left(\dfrac{\beta_\mathrm{R}\Gamma}{\Omega_{\mathrm{R}}^{s}}\right)^2 \rho_{ee,0}}{1+A_{c\rightarrow s}^{\mathrm{T}}}.
\end{equation}
This is the current, standard approach \cite{maser_few-photon_2016,turschmann_chip-based_2017}. Corresponding expressions for the reflection in this scenario can be written
\begin{equation}\label{eq:RsNorm}
    R^s_{\mathrm{norm}}= \dfrac{\Gamma^2\beta_\mathrm{L}\beta_\mathrm{R}(\rho_{ee,0})}{\left(A_{c\rightarrow s}^{\mathrm{R}}\right)},
\end{equation}
where,
\begin{equation}
    A_{c\rightarrow s}^{\mathrm{R}} =\Gamma^2\beta_\mathrm{L}\beta_\mathrm{R}\rho_{ee}^{c}
\end{equation}
 represents the control photons that have been scattered or emitted into the reflected signal channel.

\begin{figure*}
    \centering
    \includegraphics[]{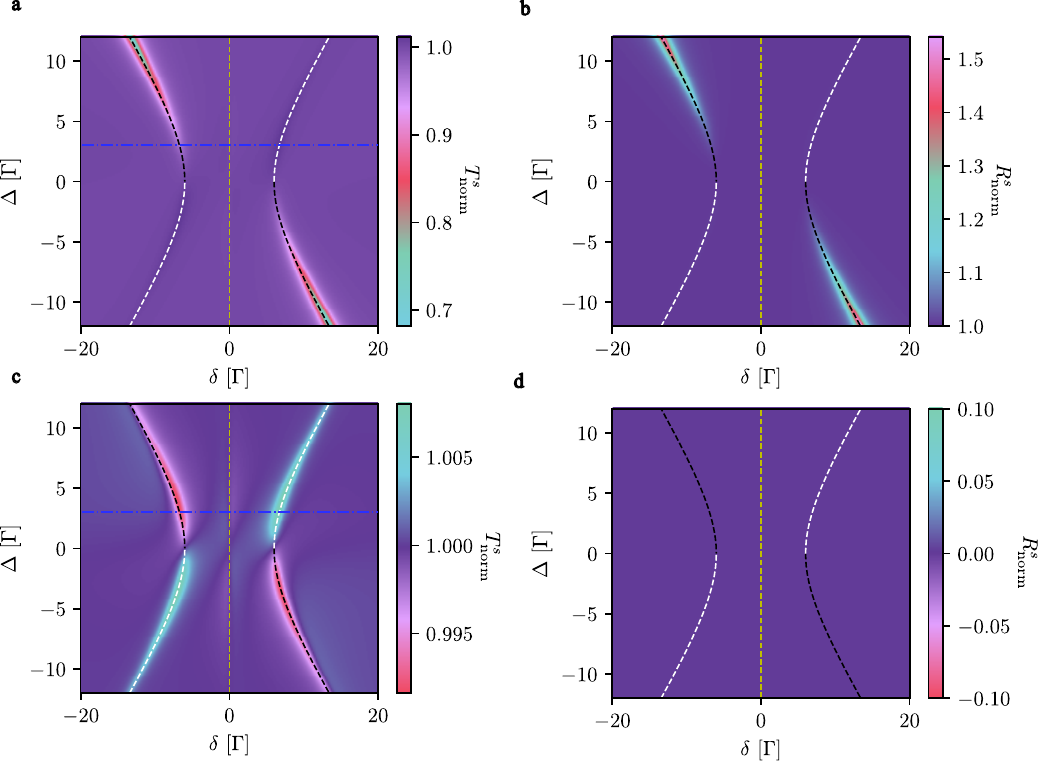}
    \caption{Frequency dependence of \(T^s_{\mathrm{norm}}\) and \(R^s_{\mathrm{norm}}\) for symmetric (a,b) and chiral (c,d) geometries. The control power is set to \(\Omega^c_{\mathrm{R}}=3\Gamma\) and signal power \(\Omega^s_{\mathrm{R}}=0.1\Gamma\). The dashed blue line is the line cut plotted below in Fig. \ref{fig:3T}.}
    \label{fig:DetMaps}
\end{figure*}

We plot the transmitted and reflected signal as a function of the detuning between the signal and control beams $\left(\delta\right)$ and control photons and emitter $\left(\Delta\right)$ for both the ideal symmetric and chiral scenarios in Fig.~\ref{fig:DetMaps}. Here, this is done for an exemplary control power $\Omega^c_\mathrm{R}=3\Gamma$. As expected in the standard dressed-state picture \cite{grynberg_central_1993, gruneisen_energy_1989, oelsner_dressed-state_2013},  we observe three distinct features in the transmission spectrum of the symmetrically-coupled TLE, marked by dashed curves in Fig.~\ref{fig:DetMaps}a. First, we observe a pronounced blue-shifted (red-shifted) extinction (black dashed curve) due to the AC-Stark effect when $\Delta<0$ $\left(\Delta>0\right)$. Second, we observe a much weaker increase in signal amplitude due to 3-photon amplification (dashed white lines) that tunes opposite to the extinction. Finally, we observe a faint signature of energy transfer between the signal and control beams at $\delta=0$ (dashed yellow line) for all $\Delta$. As expected, we also observe signatures of all three of these effects in the reflection Fig. \ref{fig:DetMaps}b, although all result in an increase of $R^s_{\mathrm{norm}}$. This explains the asymmetry between the magnitude of the extinction and amplification: a strong reflection leads to a strong extinction, while channeling new signal photons due to the amplification away from the transmission and into the reflection channel.

In contrast, there is no signature of the TLE in the reflected signal from a chirally-coupled TLE (Fig.~\ref{fig:DetMaps}d), and consequently the amplitude of the signal extinction, 0.005, is identical to that of its amplification 1.005 (Signal dominated by control emissions). In the case of amplification this is a factor of 
$\approx3$ larger than for the symmetrically-coupled emitter presented above, suggesting that coherent nonlinear quantum optics may provide a more efficient route to all-optical control of quantum light-matter interactions and quantum light states \cite{dufresne_place_nodate}. In fact, all signatures in the $T^s_{\mathrm{norm}}$ spectra arise due to energy transfer between the signal and control beams. This is discussed further in Appendix~D.

\subsubsection{The signal field alone}\label{sec:signal_only}
It is not possible to simply measure the signal field alone. As we argue above, the signal channels will always contain some coherent (i.e., scattered) and possibly incoherent (i.e., emitted) photons due to the interaction of the control beam with the TLE. Yet, in the limit where the signal field is much weaker than the control field (i.e. $\Omega_{\mathrm{R}}^{s}\ll\Omega_{\mathrm{R}}^{c}$) the effect of the signal on the population of the TLE will be small and can be treated perturbatively, meaning that,
\begin{equation}\label{eq:rho_ee_s}
    \rho^s_{ee} = \mathrm{max}\left[\rho_{ee,0}-\rho^c_{ee},|\rho_{eg,1}|^2\right],
\end{equation}
where, again, $\rho^c_{ee}$ is the emitter population measured with no signal field. Here, $|\rho_{eg,1}|^2$ represents the coherent contribution to the signal field due to $\rho^s_{ee}$ (see below) and is needed as a lower bound on this population for the cases when the addition of the signal field leads to a reduced total population (i.e., when $\rho_{ee} < \rho^c_{ee}$, due to interference effects).

Eq.~\ref{eq:rho_ee_s} allows us to write the signal (only) transmission and reflections, given by Eqs.~\ref{eq:Ts} and \ref{eq:Rs}, respectively, as,
\begin{eqnarray}
     T^{s} &\approx& 1 - \frac{2\Gamma\beta_\mathrm{R} \mathrm{Im}\left\{\rho_{eg,1}\right\}}{\Omega^s_\mathrm{R}} +\frac{\Gamma^2\beta_\mathrm{R}^2 \rho^s_{ee}}{(\Omega^s_\mathrm{R})^2}, \label{eq:Ts_only} \\
     R^s &\approx& \frac{\Gamma^2\beta_\mathrm{R}\beta_\mathrm{L} \rho^s_{ee}}{(\Omega^s_\mathrm{R})^2}. \label{eq:Rs_only}
\end{eqnarray}
The equivalent equations for the control beam are presented in Appendix D.

In the presence of noise or high photon fluxes the TLE response may not be entirely coherent, meaning that some photons are scattered or emitted with no definite phase relation to the incident signal \cite{maser_few-photon_2016}. In such a case the total signal field, in either the transmission or reflection ports, is a mixed state. Experimentally, this corresponds to filtering out and detecting only the photons in a narrow-band around $\omega_s$.

Because we are only picking out this $\omega_s$ component of the beams, the transmission and reflection can be found using the amplitude square of Eqs.~\ref{eq:ts} and \ref{eq:rs}, respectively, resulting in,
\begin{eqnarray}
    T^{s}_{\mathrm{coh}} &=& 1 - \frac{2\Gamma\beta_\mathrm{R} \mathrm{Im}\left\{\rho_{eg,1}\right\}}{\Omega^s_\mathrm{R}} +\frac{\Gamma^2\beta_\mathrm{R}^2 |\rho_{eg,1}|^2}{(\Omega^s_\mathrm{R})^2}, \label{eq:Ts_coh} \\
    R^{s}_{\mathrm{coh}} &=& \frac{\Gamma^2\beta_\mathrm{L}\beta_\mathrm{R}|\rho_{eg,1}|^2}{(\Omega^s_\mathrm{R})^2}.
\end{eqnarray}
The incoherent components can be found by taking the difference between the respective equations. For example, $T^s_\mathrm{inc}=T^s-T^s_\mathrm{coh}$.

\begin{figure}
    \centering
    \includegraphics[]{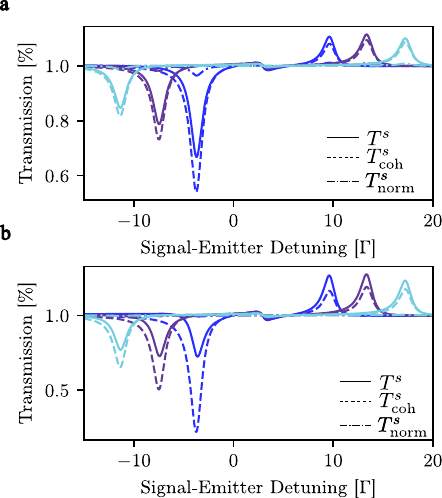}
    \caption{Line cut spectra of \(T^s\) (solid), \(T^s_{\mathrm{coh}}\) (dashed), and \(T^s_{\mathrm{norm}}\) (dot dashed) with \(\Omega^c_{\mathrm{R}}=3\Gamma,5\Gamma,7\Gamma\) (blue, purple, teal), \(\Delta=3\Gamma\), and \(\Omega^s_{\mathrm{R}}=0.1\Gamma\) for a) symmetric geometry and b) chiral geometry.}
    \label{fig:3T}
\end{figure}

The difference between the 3 transmitted signals, $T^s_{\mathrm{norm}}$ (Eq.~\ref{eq:TsNorm}), $T^s$ (Eq.~\ref{eq:Ts_only}) and $T^s_{\mathrm{coh}}$ (Eq.~\ref{eq:Ts_coh}) can be seen in Fig.~\ref{fig:3T} for the chiral and symmetric scenarios, represented by the dash-dotted, solid, and dashed lines respectively.

\begin{figure}
    \centering
    \includegraphics[]{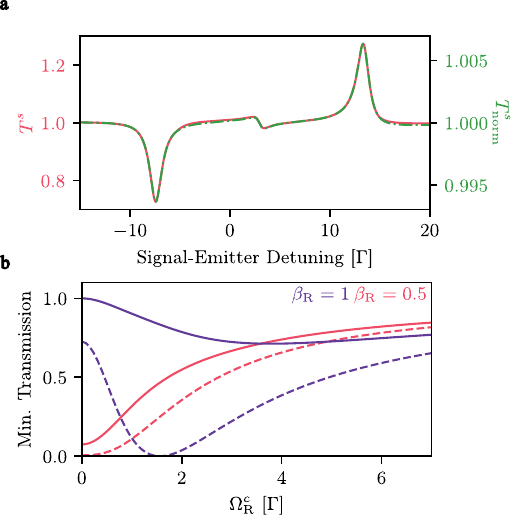}
    \caption{a) line cut spectra of $T^s_{\mathrm{norm}}$ and $T^s$ given $\Omega^c_{\mathrm{R}}=5\Gamma$, $\Delta=3\Gamma$ and $\Omega^s_{\mathrm{R}}=0.1\Gamma$. b) Shows the minimum transmission of $T^s$ (solid) and $T^s_{\mathrm{coh}}$ (dashed) across \(\omega_s\) values as $\Omega^c_{\mathrm{R}}$ in increased for $\Delta=3\Gamma$ and $\Omega^s_{\mathrm{R}}=0.1\Gamma$.}
    \label{fig:TvsCoh}
\end{figure}

The blue line in Fig.~\ref{fig:3T} corresponds to the dashed blue line on Fig.~\ref{fig:DetMaps}. Importantly, the features of \(T^s_{\mathrm{norm}}\) are only clear in the symmetric geometry at the lowest control power shown. Since \(\Omega^c_{\mathrm{R}}>>\Omega^s_{\mathrm{R}}\),the photons from the control excited emitter drown out the dynamics in the signal beam. In \(T^s\) and \(T^s_{\mathrm{coh}}\) this background is suppressed allowing for clear photon dynamics. In both the chiral and symmetric geometries, $T^s$ obeys the same behaviors as $T^s_{\mathrm{norm}}$ but are stronger from the suppressed background as highlighted in Fig.~\ref{fig:TvsCoh}a. With $T^s$ there is $\pm30\%$ in transmission in chiral system instead of $\pm0.5\%$ of $T^s_{\mathrm{norm}}$ shown in Fig.~\ref{fig:DetMaps} c. $T^s_{\mathrm{coh}}$ however, differs from the other schemes as shown in Fig. \ref{fig:TvsCoh}b. In the case $\Omega^s_{\mathrm{R}}\longrightarrow0\Gamma$ we would obtain $T^s=T^s_{\mathrm{coh}}$ when $\Omega^c_\mathrm{R}=0\Gamma$. Notably, in symmetric geometries, increasing control power also increases transmission. While in chiral geometries, the transmission decreases to a critical point, even reaching 0 for \(T^s_{\mathrm{coh}}\), before increasing. Destructive interference of the coherent components following the interaction with the TLE forces the transmitted signal photons into an incoherent regime. The behavior of $T^s_{\mathrm{coh}}$ is discussed further in Sec. IV D: An imperfect emitter.

\subsection{Phase}
Up to now, we have discussed the amplitude of the various output fields; here, we turn to the light's phase, which can be calculated from the various transmission or reflection coefficients (Eqs.~\ref{eq:tsc} and \ref{eq:rsc}). The phase shifts induced by the interaction with the TLE, relative to the respective input fields, are therefore,
\begin{subequations}\label{eq:phase}
    \begin{align}
        \Delta\varphi_t^\ell &= \mathrm{arg}\left\{t^\ell\right\},\label{phi_t} \\
        \Delta\varphi_r^\ell &= \mathrm{arg}\left\{r^\ell\right\},\label{phi_r}
    \end{align}
\end{subequations}
where $\ell = s,c$. These, we note, allow us to directly calculate the phase of the signal or control photons individually. For the signal field, for example, these would be,
\begin{subequations}\label{eq:phase_s}
    \begin{align}
        \Delta\varphi_t^s &= \mathrm{arg}\left\{1+\frac{i\Gamma\beta_\mathrm{R} \rho_{eg,1} }{\Omega_{\mathrm{R}}^{s}}\right\},\label{phi_st} \\
        \Delta\varphi_r^s &= \mathrm{arg}\left\{\frac{i\Gamma
        \sqrt{\beta_\mathrm{R}\beta_\mathrm{L}} \rho_{eg,1} }{\Omega_{\mathrm{R}}^{s}}\right\}.\label{phi_r}
    \end{align}
\end{subequations}

\begin{figure}
    \centering
    \includegraphics[width=\linewidth]{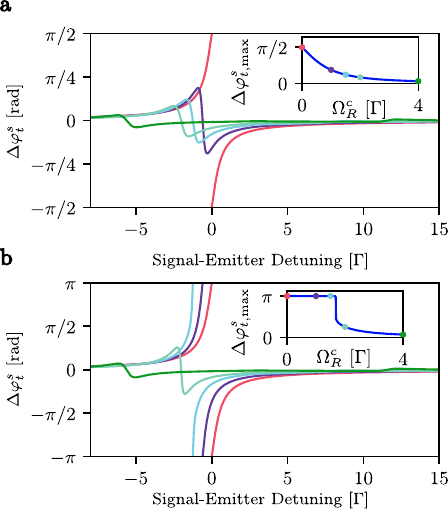}
    \caption{Signal phase as the signal frequency is swept over emitter resonance in the a) symmetric and b) chiral geometries. The control detuning is fixed (\(\Delta = 3\Gamma\)) and the control power is swept \(\Omega^c_{\mathrm{R}} =0\Gamma,1\Gamma,1.5\Gamma,2\Gamma,4 \Gamma \) denoted by red, purple, teal, mint, and green respectively for weak signal \(\Omega^s_{\mathrm{R}}=0.001\Gamma\). The insets show the maximum attainable phase as the control power is increased.}
    \label{fig:phase}
\end{figure}

We plot the phase shift of the transmitted signal field for the symmetric and chiral coupling configurations in Fig. \ref{fig:phase} a  and b, respectively, again holding $\Delta=3\Gamma$ and varying the control power. In contrast to the transmitted amplitude (Fig. \ref{fig:3T}), only the AC-Stark shift of the phase change is clearly visible, although a slight phase-shift accompanies three-photon amplification for higher powers, as is seen for signal-emitter detuning near $13\Gamma$. In terms of magnitude, the signal phase-shift induced by a symmetric-coupled TLE peaks at $\pi/2$ with no control, and rapidly decreases as $\Omega_\mathrm{R}^c$ increases (inset to Fig.~\ref{fig:phase} a). In contrast, the peak phase-shift of a chirally-coupled TLE peaks at $\pi$ and remains there until $\Omega_\mathrm{R}^c \approx 2\Gamma$, when it rapidly drops as shown in the inset to Fig.~\ref{fig:phase}b. In both cases, the peak clearly blue-shifts as the control power increases.

\subsection{An imperfect emitter}
\begin{figure*}[hbt!]
    \centering
    \includegraphics[]{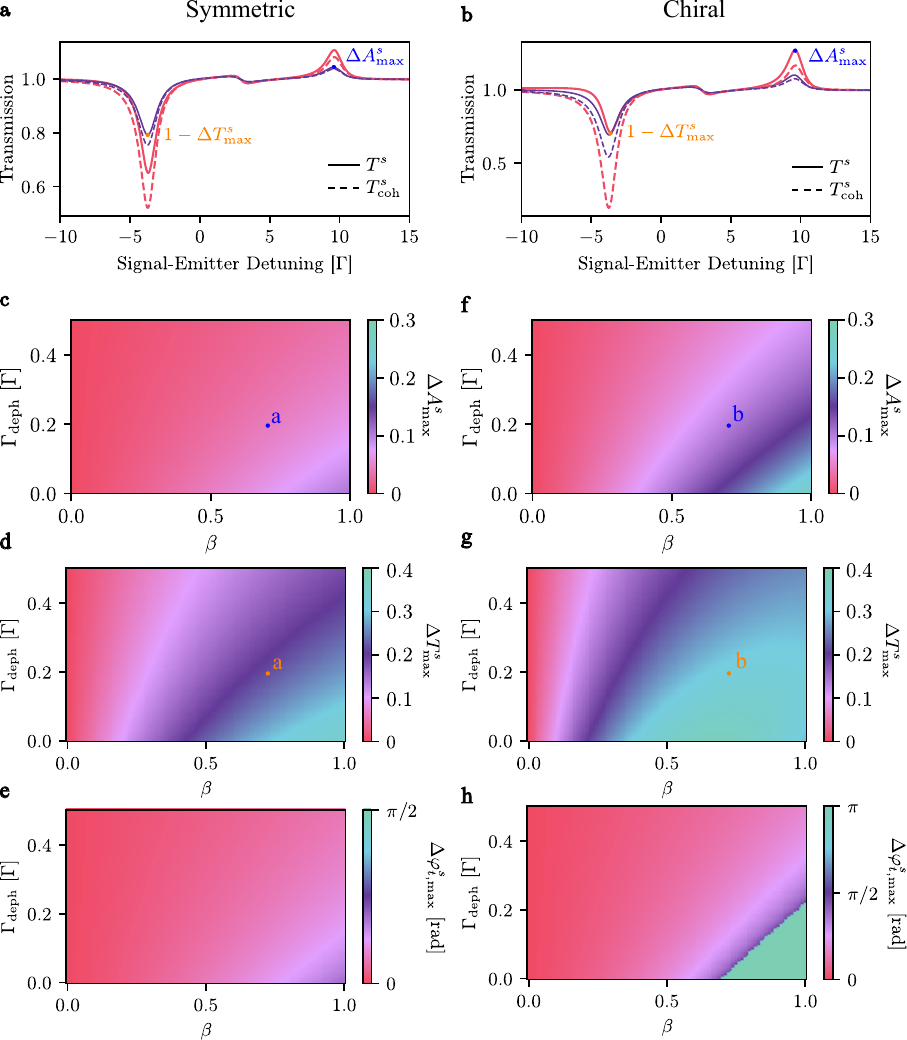}
    \caption{Line cut spectra of \(T^s\) and \(T^s_{\mathrm{coh}}\) for perfect emitter (red) and imperfect emitters (purple) with \(\beta = 0.7\) and \(\Gamma_{deph} = 0.2\Gamma\) for a) symmetric coupling and b) chiral coupling with \(\Omega^c_{\mathrm{R}}=3\Gamma\), \(\Delta=3\Gamma\), \(\Omega^s_{\mathrm{R}}=0.01\Gamma\). c)-d) and f)-g) show the impact of imperfections on the maximum amplification and extinction for symmetric and chiral coupling respectively. The corresponding points are marked in a) and b) and in c)-d)/f)-g). e) and h) demonstrate the impact of imperfections on the maximum signal phase with the same control parameters but powered down signal (\(\Omega^s_\mathrm{R}=0.001\Gamma\)).}
    \label{fig:Imperfect}
\end{figure*}

Up to now, we have presented results for ideal TLEs, meaning those with perfect coupling $\beta=1$ and no noise $\Gamma_\mathrm{deph}=0$, yet our model can account for these imperfections. We present exemplary plots of $T^s$ (solid curve) and $T^s_\mathrm{coh}$ (dashed curve) that demonstrate the consequence of imperfect TLEs in Fig.~\ref{fig:Imperfect}. First, Fig.~\ref{fig:Imperfect}a shows the symmetric case when $\Delta=3\Gamma$ and $\Omega_\mathrm{R}^c=3\Gamma$, for the ideal scenario (red), when losses are introduced ($\beta=0.7$) and when noise is present $\Gamma_\mathrm{deph}=0.2\Gamma$ (purple). As expected, both imperfections reduce the peak extinction $\left(\Delta T_\mathrm{max}\right)$ of the Stark-shifted resonance (top panel), while pure-dephasing also causes it to broaden. Note, too, that the addition of dephasing reduces the coherent component of the scattered field and the phase-shift that this component experiences (bottom panel). Reduction in the coherent component of the scattered field decreases the destructive interference with the incident field which results in an increased \(T^s_{\mathrm{coh}}\). Similarly, introducing imperfections reduced the peak coherent amplification $\left(\Delta A_\mathrm{max}\right)$.

We observe a similar behavior for a chirally-coupled imperfect TLE, shown in Fig.~\ref{fig:Imperfect}b, although the details differ. This is most notable in the peak achievable phase change, which remain $\Delta\varphi_\mathrm{max}=\pi$ for the chirally coupled emitter when $\Gamma_\mathrm{deph}=0.2\Gamma$ (\(\beta=1\)) while for the symmetric case $\Delta\varphi_\mathrm{max}$ drops to $\approx\pi/8$ from the ideal $\Delta\varphi_\mathrm{max}=\pi/2$.We summarize the dependence of $\Delta T_\mathrm{max}$, $\Delta A_\mathrm{max}$ and $\Delta\varphi_\mathrm{max}$ on $\beta$ and $\Gamma_\mathrm{deph}$ for the symmetric and chirally coupled TLEs in Fig.~\ref{fig:Imperfect} c-e and f-h, respectively.

Although \(T^s\) always remains larger in the chiral case, this is no longer true when looking at \(T^s_{\mathrm{coh}}\). The extinction in the chiral case reaches \(\approx\)80\%, whereas the symmetric geometry reaches \(\approx\)45\%. The destructive interference of the coherent scattered field leads to reduced coherence in the transmitted field. Chiral geometries show more transmission but less coherent photons. For all quantities studied (\(\Delta A^s_{\mathrm{max}}\), \(\Delta T^s_{\mathrm{max}}\), and \(\Delta \varphi^s_{\mathrm{t,max}}\)) chiral coupling exhibits clear robustness to imperfections when compared to symmetric coupling.

\section{Conclusions}
We have presented a model for coherent multicolor nonlinear optics with TLEs in a one dimensional waveguide, both in the symmetric and chiral configurations. Just as in the single color case, the total transmission remains untouched, leaving the interaction imprinted on the phase. However, looking deeper, we discover a rich world of photon transfer, frequency conversion, and complex dynamics between the signal and control beams. Using the methods presented we can efficiently filter out the control beam and uncover the true strength of the signal photon dynamics, increasing signatures by x100.
We have displayed how, with chiral geometries, we can achieve an optically tunable $\pi$ phase shift with strong robustness to dephasing and poor coupling, opening doors to future all optical switch for photon routing in quantum networks.

\section{Acknowledgments}
The authors thank John E. Sipe for valuable discussions on the quantum theory, and gratefully acknowledge the support from the National Research Council of Canada (NRC), the Canadian Foundation for Innovation (CFI), the Ontario Ministry of Colleges, Universities, Research Excellence and Security (MCURES), the Natural Sciences and Engineering Research Council of Canada (NSERC), and Queen's University.

\clearpage
\bibliography{chiral_nonlinear_optics_and_optical_control}
\clearpage
\appendix
\counterwithin{figure}{section}
\section{Density Matrix elements and Truncation}
\begin{figure}[htp]
    \centering
    \includegraphics[width=\linewidth]{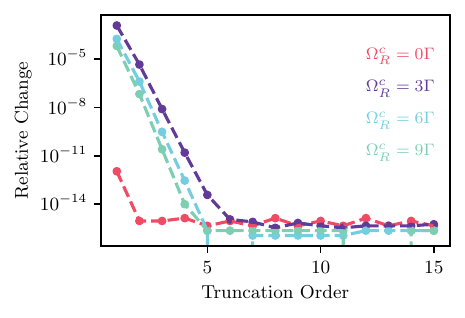}
    \caption{Truncation analysis for increasing control power and \(\Omega^s_{\mathrm{R}}=0.1\Gamma\) and \(\Delta=3\Gamma\) calculated for \(T^s_{\mathrm{coh}}\).}
    \label{trunc}
\end{figure}

Truncation analysis is conducted by calculating the maximum error between signal sweeps of observables for \(\mathrm{trunc}=n\) and \(\mathrm{trunc}=n-1\). Fig. A.1 shows an exemplary truncation analysis for \(T^s_\mathrm{coh}\). A truncation order of \(n=10\) was used for figures shown.   

From equations \ref{eq:rho_ee_n} and \ref{eq:rho_ge_n} we have expressions for the density matrix elements in the rotating frame at the control frequency as
\begin{eqnarray}
    \rho_{ee}&=&\sum\rho_{ee,n}e^{in\delta t},\rho_{ee,n}=w_{n}+\frac{1}{2}\delta_{n0}\label{eq:rho_eeApp}
    \\
   \rho_{ge}&=&\sum\rho_{ge,n}e^{in\delta t},\rho_{ge,n}=u_{n}+iv_{n} \label{eq:rho_geApp}.
\end{eqnarray}

When considering observables we are interested in the time average expectation value. Both transmission and phase calculations depend on coherent interactions between incident fields and the scattered field. Each matrix elements has oscillatory behavior in harmonics incremented by the beating frequency of the input fields. For the time average, it is required for \(e^{-in\delta t}\) to equal 0 to remove oscillations. Of note, for signal-scattered field interaction, the beating frequency add an additional \(e^{-i\delta t}\). Thus, unlike for the control (eq. \ref{eq:avgcont_app}) the signal interaction corresponds to \(n=1\) (eq. \ref{eq:avgsig_app}) and not \(n=0\). 
\begin{eqnarray}
    \left\langle{\rho}_{ge}e^{-i\delta t}\right\rangle &=& \left\langle\  \sum {\rho}_{ge,n} e^{-i(n-1)\delta t} \right\rangle ={\rho}_{ge,1}\label{eq:avgsig_app}
    \\
    \left\langle\rho_{ge}\right\rangle &= &\left\langle\  \sum {\rho}_{ge,n} e^{-in\delta t} \right\rangle ={\rho}_{ge,0}\label{eq:avgcont_app}
    \\
    \left\langle\rho_{ee}\right\rangle &=& \left\langle\  \sum {\rho}_{ee,n} e^{-in\delta t} \right\rangle ={\rho}_{ee,0}\label{eq:avgpop_app}
\end{eqnarray}

\section{Scattered Fields}
First we consider the form of the electric fields in a waveguide following the Green's Tensor approach \cite{asenjo-garcia_atom-light_2017} where we generalize the expression to account for the signal and control both left and right propagating. Where $\hat{\boldsymbol{f}}$ denotes the bosonic operator of the fields and  $\boldsymbol{G}$ the Green's Tensor.
\begin{widetext}
    \begin{eqnarray}
            \hat{\boldsymbol{E}}_{d}^{\ell+}(r,\omega_s)=i\mu\omega_\ell^{2}\sqrt{\frac{\hbar\epsilon_{o}}{\pi}}\int d\boldsymbol{r'}\sqrt{\epsilon_{I}(\boldsymbol{r'},\omega_s)}\boldsymbol{G}^\ell(\boldsymbol{r,}\boldsymbol{r'},\omega_\ell)\cdot\hat{\boldsymbol{f}}^\ell_{d}(\boldsymbol{r'},\omega_\ell)\\
    \end{eqnarray}
\end{widetext}
where, recall, $d=\mathrm{L,R}$ and $\ell=s,c$.
We look at the time evolution of the bosonic operator using the Von Neumann equation
\begin{equation}
    \frac{d}{dt}\hat{\boldsymbol{f}}_{i}(\boldsymbol{r},\omega_i)=\frac{i}{\hbar}\left[\hat{\boldsymbol{H}_{s}},\hat{\boldsymbol{f}}_{i}(\boldsymbol{r},\omega_i)\right]
\end{equation}
following the commutation relation,
\begin{equation}
    \left[\hat{\boldsymbol{f}}^\ell
    _{d}(\boldsymbol{r},\omega_\ell),\hat{\boldsymbol{f}}_{d'}^{\ell'\dagger}(\boldsymbol{r},\omega_{\ell'})\right]=\delta_{d,d'}\delta(r-r')\delta_{d,d'}(\omega_\ell-\omega_{\ell'})
\end{equation}
Computing the commutation results in 

\begin{widetext}
    \begin{eqnarray}
        \frac{d}{dt}\hat{\boldsymbol{f}}^\ell_{d}(\boldsymbol{r},\omega_d)=-i\omega_d\hat{\boldsymbol{f}}^\ell_{d}(\boldsymbol{r},\omega_d)+\mu\omega_d^{2}\sqrt{\frac{\hbar\epsilon_{o}}{\pi}}\boldsymbol{d}(\boldsymbol{r_{A}})\sqrt{\epsilon_{I}(\boldsymbol{r},\omega_d)}\boldsymbol{G}^{\ell*}_{d}(\boldsymbol{r_{A},}\boldsymbol{r},\omega_d)\hat{\boldsymbol{\sigma}}_{ge}e^{i\omega_d t}\\
    \end{eqnarray}
\end{widetext}
We integrate from time t to t' which results in a total expression for the field following interaction with the quantum emitter.

\begin{widetext}
    \begin{equation}
    \hat{\boldsymbol{f}}^{\ell}_d(\boldsymbol{r},\omega_k)=\hat{\boldsymbol{f}}^\ell_{d,free}(\boldsymbol{r},\omega_\ell)e^{-i\omega_\ell(t-t')}+\mu\omega_\ell^{2}\sqrt{\frac{\hbar\epsilon_{o}}{\pi}}\boldsymbol{d}(\boldsymbol{r_{A}})\sqrt{\epsilon_{I}(\boldsymbol{r},\omega_\ell)}\boldsymbol{G}^{\ell*}(\boldsymbol{r_{A},}\boldsymbol{r},\omega_\ell)\hat{\boldsymbol{\sigma}}_{ge}e^{-i\omega_\ell(t-t')}
    \end{equation}
\end{widetext}
Here we clearly see that the field after the emitter is the sum of the incident field and the scattered field generated by the atomic operator $\hat{\boldsymbol{\sigma}}_{ge}$ such that we can express the total field as

\begin{eqnarray}
    \hat{\boldsymbol{E}}^{\ell+}(r,\omega_\ell) = \hat{\boldsymbol{E}}_{I}^{\ell+}(r,\omega_\ell) + \hat{\boldsymbol{E}}_{S}^{\ell+}(r,\omega_\ell).
\end{eqnarray}

We simplify our expression for the scattered field by following \cite{asenjo-garcia_atom-light_2017} such that
\begin{eqnarray} \label{eq:43}
    \hat{\boldsymbol{E}}_{S}^{+}(r,t)=\mu_{o}\omega_A^{2}\boldsymbol{G_{k}}(\boldsymbol{r,}\boldsymbol{r_{A}},\omega_A)\cdot\boldsymbol{d}\hat{\boldsymbol{\sigma}}_{ge}(t)
\end{eqnarray}

This makes up the total scattered field we can see this by looking at the expectation value,

\begin{eqnarray}
    \langle\hat{\boldsymbol{E}}_{S}^{+}(r,t) \rangle\sim \langle\hat{\boldsymbol{\sigma}}_{ge}(t)\rangle \sim \sum \rho_{eg,n}e^{-in\delta t}
\end{eqnarray}

Where we see clearly that the atomic operator is generating fields at the signal and control frequency as well in harmonics separated by $\delta$ from 4-wave mixing. Thus this represents the \textbf{total} scattered field in 1 term.
\section{Total Transmission and Reflection}

Now we can shift our focus towards transmission which is given by

\begin{equation}
    T=\frac{\left\langle \hat{E}^{-}\hat{E}^{+}\right\rangle }{\left\langle \hat{E}_{inc}^{-}\hat{E}_{inc}^{+}\right\rangle }
\end{equation}
In which we look at the transmitted intensity of light normalized by the incident intensity. Clearly we have

\begin{equation}
    \hat{E}^{-}=\hat{E}_{R}^{s-}+\hat{E}_{R}^{c-}+\hat{E}_{R}^{scat-}
\end{equation}

and that the incoming fields are the signal and control beams. A measurement is essentially acting with the dipole \(\textbf{d}^*\cdot\) which then allows us to write the scattered field in terms of \(\beta\) from the dipole projected Green's Tensor. The transmitted field then becomes

\begin{widetext}
\begin{equation}
T = \left\langle \left(\boldsymbol{d^{*}}\cdot\hat{E}_{R}^{s-}+\boldsymbol{d^{*}}\cdot\hat{E}_{R}^{c-}-i\Gamma\beta\hat{\sigma}_{eg}f^*(r)\right)\left(\boldsymbol{d^{*}}\cdot\hat{E}_{R}^{s+}+\boldsymbol{d^{*}}\cdot\hat{E}_{R}^{c+}+i\Gamma\beta\hat{\sigma}_{ge}f(r)\right)\right\rangle 
\end{equation}
\end{widetext}

in which \(f(r)\) contains the spatial evolution of the scatter field which is simply \(e^{ik_{\ell}}\) for the component at frequency \(\omega_\ell\). To treat \(\boldsymbol{d^{*}}\cdot\hat{E}_{R}^{s-}\) we remember our definition of the rabi frequency which occurs at the emitter \((\hat{\Omega}=\hat{d}^{*}\cdot \hat{E}^{+}(r = 0))/\hbar\) we can express our incident fields as \(\hat{E}_{R}^{s-}=\hat{E}_{R}^{s-}(r=0)e^{i(k_{s}r+i\omega_{s}t)}\).

Expanding out our expression for the unnormalized transimission we get 

\begin{widetext}
\begin{equation} 
\begin{aligned}
    T =  \left\langle \hat{\Omega}_{R}^{s}\right\rangle ^{2}+\left\langle \hat{\Omega}_{R}^{c}\right\rangle ^{2}+\left\langle \hat{\Omega}_{R}^{s}\hat{\Omega}_{R}^{c}e^{i\delta t}\right\rangle -2Im\{\beta_{R}\Gamma\Omega_{R}^{s}\left\langle \hat{\sigma}_{ge}e^{i\omega_{s}t}\right\rangle \}-2Im\{\beta_{R}\Gamma\Omega_{R}^{c}\left\langle \hat{\sigma}_{ge}e^{i\omega_{c}t}\right\rangle \}+\left(\beta_{R}\Gamma\right)^{2}\left\langle \hat{\sigma}_{eg}\hat{\sigma}_{ge}\right\rangle 
    \end{aligned}
\end{equation}
\end{widetext}

In which we can see the terms represent the signal field, control field, the interaction between the input fields. The following two terms are the incident fields that scatter off the emitter, interacting destructively and the last term are the photons absorbed and then emitted by the emitter and contains emissions at both the signal and control frequency. Here, the cross terms (i.e. interactions between different frequency fields) have their time average response go to zero such as the third term. The detuning between the fields will cause oscillations at \(\delta\) will cause variation in time for constructive or destructive interference resulting in an expectation value of 0 and thus does not appear in time independent transmission values. Beating would likely be observed when taking time-resolved measurements as well as potential four-wave mixing.

\begin{equation}
    T_{norm} =  \left\langle \hat{\Omega}_{R}^{s}\right\rangle ^{2}+\left\langle \hat{\Omega}_{R}^{c}\right\rangle ^{2}
\end{equation}

Inserting our expectation values for the atomic operators shown in section A1 and noting that the expectation value of the incident field interactions will go to zero the full transmission equation is
\begin{widetext}
\begin{eqnarray}
    T=\frac{1}{(\Omega_{R}^{s})^{2}+(\Omega_{R}^{c})^{2}}\left((\Omega_{R}^{s})^{2}+(\Omega_{R}^{c})^{2}-2Im\{\Omega_{R}^{s}\beta_{R}\Gamma\rho_{eg,1}\}-2Im\{\Omega_{R}^{c}\beta_{R}\Gamma\rho_{eg,0}\}+\left(\beta_{R}\Gamma\right)^{2}\rho_{ee,0}\right).
\end{eqnarray}
\end{widetext}
\section{Control Field}

In section B we present an expression for $T^s_{\mathrm{norm}}$. The missing piece is what happens to the control beam following the interaction. We will be left with the control field and its interaction with the scattered field at the control frequency. In this case we obtain,

\begin{eqnarray}
    T^c_{\mathrm{norm}} = 1 - \frac{2\Gamma\beta_\mathrm{R}\mathrm{Im}\left\{\rho_{eg,0}\right\}}{(\Omega^c_\mathrm{R})}
\end{eqnarray}

\begin{figure}
    \centering
    \includegraphics[]{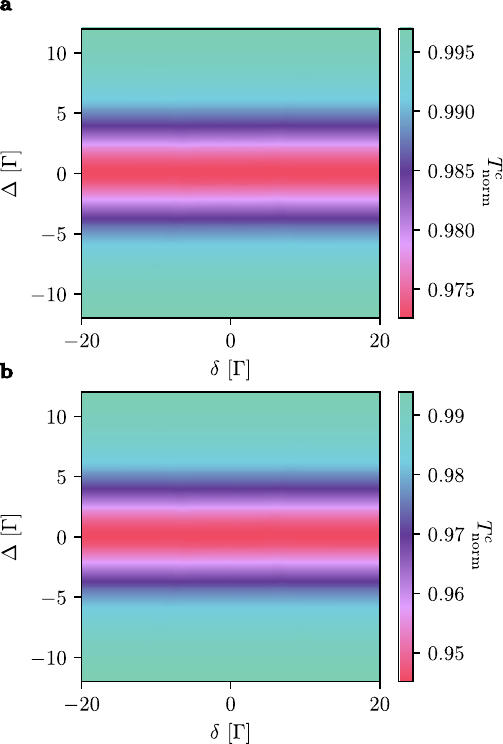}
    \caption{Exemplary $T^c_{\mathrm{norm}}$ figure for symmetric and chiral geometries respectively with $\Omega^c_{\mathrm{R}} = 3\Gamma$, $\Omega^s_{\mathrm{R}}=0.1\Gamma$, $\beta=1$, and $\Gamma_{\mathrm{deph}}=0\Gamma$. }
    \label{fig:Tcnorm}
\end{figure}

In this scheme, the scattered photons are directed in the signal path. Fig. \ref{fig:Tcnorm}a and b shows $T^c_{\mathrm{norm}}$ in the symmetric and chiral cases respectively. The greater coupling efficiency in chiral geometries results in more scattered and emitted photons by the emitter. Thus more photons are taken from the control and scattered into the signal path. The difference in extinction mirrors the difference in coupling; twice the coupling, twice the extinction.

In section B: The signal field alone, we present approximate equations for \textit{only} the transmitted and reflected signal fields. Here we present the corresponding formulas for the control field. The equations have similar form to the signal giving,

\begin{eqnarray}
     T^{c} &\approx& 1 - \frac{2\Gamma\beta_\mathrm{R} \mathrm{Im}\left\{\rho_{eg,0}\right\}}{\Omega^c_\mathrm{R}} +\frac{\Gamma^2\beta_\mathrm{R}^2 \rho^c_{ee}}{(\Omega^c_\mathrm{R})^2}, \label{eq:Tc_only} \\
     R^c &\approx& \frac{\Gamma^2\beta_\mathrm{R}\beta_\mathrm{L} \rho^c_{ee}}{(\Omega^c_\mathrm{R})^2}. \label{eq:Rc_only}
\end{eqnarray}

where the terms in eq. \ref{eq:Tc_only} correspond to the control field, the control-scattered field interaction, and the scattered control field. To extract the coherent control transmission we perform the same trick used for eq. \ref{eq:Ts_coh}. Thus, the coherent control can be written as,

\begin{eqnarray}
     T_{\mathrm{coh}}^{c} &=& 1 - \frac{2\Gamma\beta_\mathrm{R} \mathrm{Im}\left\{\rho_{eg,0}\right\}}{\Omega^c_\mathrm{R}} +\frac{\Gamma^2\beta_\mathrm{R}^2 |\rho_{eg,0}|^2}{(\Omega^c_\mathrm{R})^2}, \label{eq:Tc_coh} \\
\end{eqnarray}

\begin{figure}
    \centering
    \includegraphics[]{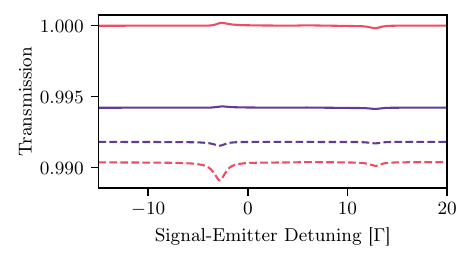}
    \caption{$T^c$ (solid) and $T^c_{\mathrm{coh}}$ (dashed) for chiral (red) and symmetric (purple) geometries. With $\Omega^c_{\mathrm{R}} = 3\Gamma$, $\Omega^s_{\mathrm{R}}=0.1\Gamma$, $\beta=1$, and $\Gamma_{\mathrm{deph}}=0\Gamma$. }
    \label{fig:controlT}
\end{figure}

An important distinction is that these are normalized to the control beam power. In the perfect chiral case with directionality 1, all changes in transmission is a result of photon transfer between fields. Since $\Omega^c_{\mathrm{R}}>>\Omega^s_{\mathrm{R}}$ in typical experiments, the features in the control beam are significantly drowned out by the control power. An important note is that the reflected control field in this model does not change as the signal interacts with the emitter but the change is expected to be minor. In transmission, the change in control is on the order $\sim0.01\%$. As shown in Fig. \ref{fig:controlT} the total and coherent control transmission only vary by less than a percent as most of the control photons are not interacting at this power.

\end{document}